\documentclass[conference]{IEEEtran}
\IEEEoverridecommandlockouts
\usepackage{cite}
\usepackage{amsmath,amssymb,amsfonts}
\usepackage{algorithmic}
\usepackage{graphicx}
\usepackage{textcomp}
\usepackage{xcolor}
\def\BibTeX{{\rm B\kern-.05em{\sc i\kern-.025em b}\kern-.08em
    T\kern-.1667em\lower.7ex\hbox{E}\kern-.125emX}}
    
\usepackage{enumitem}
 
\makeatletter
\def\ps@IEEEtitlepagestyle{%
  \def\@oddfoot{\mycopyrightnotice}%
  \def\@evenfoot{}%
}
\def\mycopyrightnotice{%
  {\footnotesize This work has been submitted to the IEEE for possible publication. Copyright may be transferred without notice, after which this version may no longer be accessible.\hfill}% <--- Change here
  \gdef\mycopyrightnotice{}% just in case
}  
    
\begin{document}
%%%%%%%%%%%%%%%%%%%%%%%%%%%%%%
%TITLE
%%%%%%%%%%%%%%%%%%%%%%%%%%%%%%
\title{A Natural Language Interface with Relayed Acoustic Communications for Improved Command and Control of AUVs\\
%{\footnotesize \textsuperscript{*}Note: Sub-titles are not captured in Xplore and should not be used}
%\thanks{EPSRC ORCA Hub, UK MOD Dstl, RAEng/Leverhulme Trust.}
%HH commented  out as strange
%DR This is part of the IEEE template
}

%%%%%%%%%%%%%%%%%%%%%%%%%%%%%%
%AUTHORS
%%%%%%%%%%%%%%%%%%%%%%%%%%%%%%

%I am using an alternative style of author block (i.e. different to that in the trans template. It  matches the style I have observed in AUV2016 papers

%I quote the IEEE trans template guidance "Please keep your affiliations as succinct as possible (for example, do not differentiate among departments of the same organization)."

%DR - Maybe the PIs can confirm the author list and order please?
\author{\IEEEauthorblockN{David A. Robb\IEEEauthorrefmark{2}, Jonatan Scharff Willners\IEEEauthorrefmark{1}\IEEEauthorrefmark{2}, Nicolas Valeyrie\IEEEauthorrefmark{2}, Francisco J. Chiyah Garcia\IEEEauthorrefmark{2}, Atanas Laskov\IEEEauthorrefmark{1} }{ Xingkun Liu\IEEEauthorrefmark{2}, Pedro Patron\IEEEauthorrefmark{1}, Helen Hastie\IEEEauthorrefmark{2}, Yvan R. Petillot\IEEEauthorrefmark{2} }

\IEEEauthorblockA{\IEEEauthorrefmark{2}Heriot-Watt University, Edinburgh, UK \{d.a.robb, js100, n.valeyrie, fjc3, x.liu, h.hastie, y.r.petillot\}@hw.ac.uk}\IEEEauthorblockA{\IEEEauthorrefmark{1}SeeByte Ltd., Edinburgh, UK, \{jonatan.willners, atanas.laskov, pedro.patron\}@seebyte.com}}

%\author{\IEEEauthorblockN{1\textsuperscript{st} Given Name %Surname}
%\IEEEauthorblockA{\textit{dept. name of organization (of Aff.)} %\\
%\textit{name of organization (of Aff.)}\\
%City, Country \\
%email address}
%\and
%\IEEEauthorblockN{2\textsuperscript{nd} Given Name Surname}
%\IEEEauthorblockA{\textit{dept. name of organization (of Aff.)} %\\
%\textit{name of organization (of Aff.)}\\
%City, Country \\
%email address}
%\and
%\IEEEauthorblockN{3\textsuperscript{rd} Given Name Surname}
%\IEEEauthorblockA{\textit{dept. name of organization (of Aff.)} %\\
%\textit{name of organization (of Aff.)}\\
%City, Country \\
%email address}
%\and
%\IEEEauthorblockN{4\textsuperscript{th} Given Name Surname}
%\IEEEauthorblockA{\textit{dept. name of organization (of Aff.)} %\\
%\textit{name of organization (of Aff.)}\\
%City, Country \\
%email address}
%\and
%\IEEEauthorblockN{5\textsuperscript{th} Given Name Surname}
%\IEEEauthorblockA{\textit{dept. name of organization (of Aff.)} %\\
%\textit{name of organization (of Aff.)}\\
%City, Country \\
%email address}
%\and
%\IEEEauthorblockN{6\textsuperscript{th} Given Name Surname}
%\IEEEauthorblockA{\textit{dept. name of organization (of Aff.)} %\\
%\textit{name of organization (of Aff.)}\\
%City, Country \\
%email address}
%}

%%%%%%%%%%%%%%%%%%%%%%%%%%%%%%
%maketitle
%%%%%%%%%%%%%%%%%%%%%%%%%%%%%%
\maketitle

%%%%%%%%%%%%%%%%%%%%%%%%%%%%%%
%ABSTRACT
%%%%%%%%%%%%%%%%%%%%%%%%%%%%%%
\begin{abstract}
Autonomous underwater vehicles (AUVs) are being tasked with increasingly complex missions. The acoustic communications required for AUVs are, by the nature of the medium, low bandwidth while adverse environmental conditions underwater often mean they are also intermittent. This has motivated development of highly autonomous systems, which can operate independently of their operators for considerable periods of time. These missions often involve multiple vehicles leading  not only to challenges in communications but also in command and control (C2). Specifically operators face complexity in controlling multi-objective, multi-vehicle missions, whilst simultaneously facing uncertainty over the current status and safety of several remote high value assets. Additionally, it may be required to perform command and control of these complex missions in a remote control room. In this paper, we propose a combination of an intuitive, natural language operator interface combined with communications that use platforms from multiple domains to relay data over different mediums and transmission modes, improving command and control of collaborative and fully autonomous missions. In trials, we have demonstrated an integrated system combining working prototypes with established commercial C2 software that enables the use of a natural language interface to monitor an AUV survey mission in an on-shore command and control centre.
%HH is "different medium modalities" correct? or is it "different mediums" or different modalities" both seems strange-- this is also used in the text
%DR I wasn't sure about this either. I have changed. @PP Does the change reflect what we wish? 
\end{abstract}

%%%%%%%%%%%%%%%%%%%%%%%%%%%%%%
%keywords
%%%%%%%%%%%%%%%%%%%%%%%%%%%%%%

%%%%%%%%%%%%%%%%%%%%%%%%%%%%%%
%Introduction
%%%%%%%%%%%%%%%%%%%%%%%%%%%%%%
\section{Introduction}
%\textbf{WE NEED TO REWRITE AND OR AUGMENT THIS FROM THE ACCEPTED ABSTRACT}

\begin{figure}[t]
\centering
\includegraphics[width=0.94\linewidth]{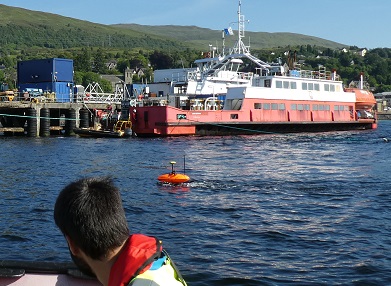}
\caption{An EvoLogics Sonobot autonomous surface vehicle (centre of frame) heading to its station ready to relay acoustic communications (ACOMMS) over WiFi. The vehicle carries an EvoLogics modem and on-board computer to receive the ACOMMS, package them and send the payload onward over WiFi.}

\label{fig:sonobot}
\end{figure}

%%JSW: I dont understand this part. UWSN is not part of this work.

%%PP Suggestion:
% The problem is sampling, processing and transfer of data
Underwater environmental conditions such as high pressure, salinity corrosion, darkness and sound attenuation make data gathering, processing and transfer underwater very difficult. % Data gathering
Underwater sensor networks (UWSN) can now be deployed to efficiently sample data from large areas over long periods of time.  These are proving to have benefits for data gathering in applications such as tracking shipping traffic, environmental monitoring, and detection of events such as earthquakes and oil spills. 

% First link for data transfer - acoustic
The use of moving Autonomous Underwater Vehicles (AUVs) to gather data from nodes in the network can increase the life-time of these networks. 
Instead of messages having to be transmitted between multiple nodes, only a close proximity transfer between the sensor node and the AUV is required.
%Communication there is limited to acoustic signals, where low bandwidth, short range and high latency is the norm.
Communication deep underwater is limited to acoustic signals, where low bandwidth, short range and high latency is the norm.
As such, there is no close remote control loop available to the operator, therefore, requiring the AUVs to perform their tasks unattended. This means they have to be able to dynamically adapt to the environment by enabling high levels of autonomy, as through the Neptune Autonomy Framework by SeeByte used in the study described here.

This means that the missions these platforms perform are becoming more complex, often involving coordination of multiple vehicles and adaptive tasking of multiple objectives.  This poses a challenge to operators, where missions can be adversely affected by operators lacking understanding of the true status of the mission, the autonomous vehicles and the reasons for their behaviour. Such lack of transparency can reduce trust and ultimately acceptance and adoption. 

%The problems of communications with submerged Autonomous Underwater Vehicles (AUVs) (low band-width and intermittent communications) have necessitated the development of AUVs with a high level of autonomy. This has enabled multiple vehicles to collaborate on missions with multiple objectives. Such complex missions present a challenge in command and control. When vehicles begin to behave contrary to the expectations of operators, a mission can be adversely affected by unnecessary decisions to recall a vehicle or abort the mission due to operators lacking understanding of the true status of the vehicles and the reasons for their behaviour.
%Autonomous vehicle missions are becoming more complex, involving multiple vehicles 
%%PP Is this the correct citation here? I can't see the relationship
%\cite{Matsebe2018}, possibly across multiple domains (e.g. underwater, surface and air), and multi-component objectives. 

%%PP 
% Second data link for data transfer - 
The increased endurance of these platforms is now allowing for missions covering longer distances and even over-the-horizon operations.
These missions are generally being monitored from a command and control (C2) centre located in a remote location such as a shore-based station or a vessel of opportunity. As a consequence, the range of the data exchanges
between the AUVs and the C2 station need to be extended. This can be achieved by deploying additional autonomous systems, such as Unmanned Surface Vehicles (USVs), to act as a communication bridge.
%bridging between the different communication modalities.

These are generally acoustic for below the surface and electromagnetic for above the surface.
This increases the complexity of the missions even further, by involving platforms operating across multiple domains (e.g. underwater, surface and air).
%It is likely such a diversity of assets will require oversight and control from a location more remote than from a nearby vessel. It might be desirable to have a distant command and control centre. 
%This would require communications relays. 
%In the case of acoustic communications (ACOMMS) from AUVs, an Autonomous Surface Vehicle (ASV) could be used to relay the communication over WiFi or other electromagnetic means of communication. 

%%PP What is the main contribution of the paper? Something like...
In this paper, we propose a combination of an intuitive natural language chat operator interface and a communication behaviour scheme that uses platforms from multiple domains to relay data over different medium modalities, while performing a collaborative and fully autonomous mission.

\begin{figure*}[t]
\centering
\frame{\includegraphics[width=0.94\linewidth]{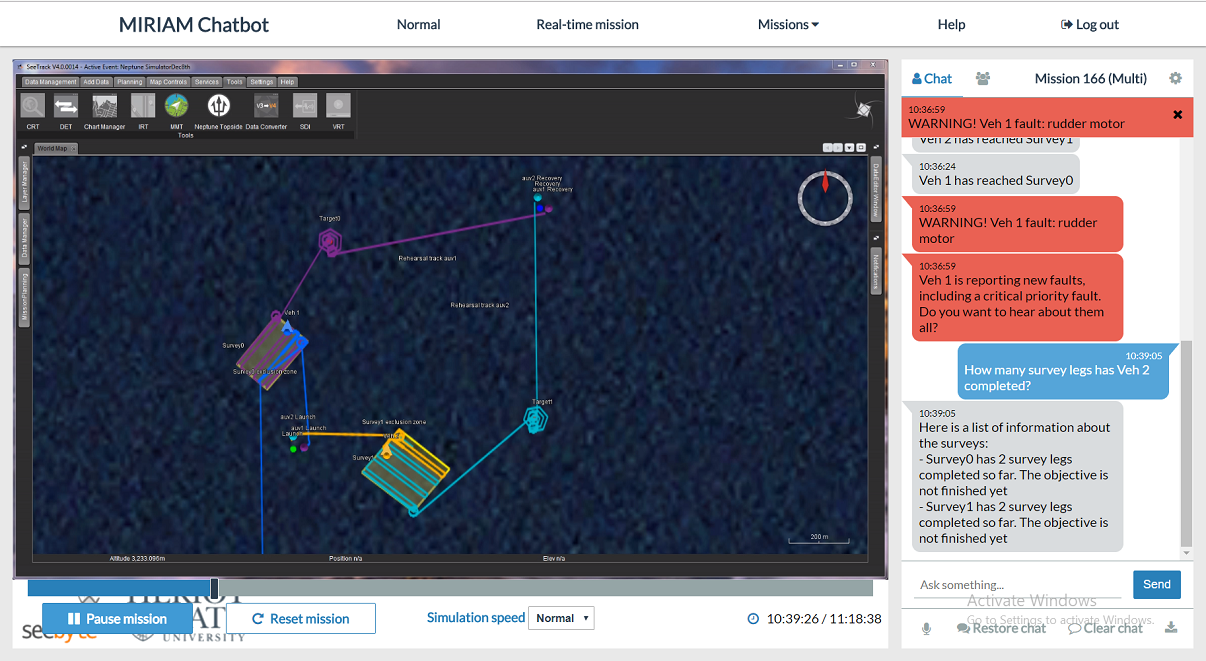}}
\caption{The MIRIAM interface combined with SeeTrack showing the predicted path of the vehicles on the left and the natural language chat on the right. This is a screen capture from the simulation application that was built for demonstration and evaluation purposes. It allows interaction with several different recorded simulated missions. This particular screen capture shows a mission and natural language interaction involving two AUVs on a mission consisting of two survey and two target requisition objectives. MIRIAM, in this case, has a pinned warning about a fault, which will persist while the relevant conversation gradually scrolls up out of view. It also shows how mission information can be queried and is displayed, with a query and answer about the progress of the current survey objectives.}
\label{fig:miriam2vehchatfilled}
\end{figure*}

%We have developed a 
%% 1) The chat interface
The natural language interface, known as MIRIAM (Multimodal Intelligent inteRaction for Autonomous systeMs), allows operators to naturally and intuitively query an autonomous system about the progress of the mission objectives and the status of the AUVs tasked to undertake it \cite{Hastie17demo}. 
%%PP I don't understand the sentence below: 'map and table?'
%Augmenting the existing map and table AUV monitoring interface with the chatbot (Figure\ref{fig:3parscreenlive}) has already been shown to improve situation awareness for experienced operators in a study with simulated AUV missions \cite{Robb2018}.
Augmenting the existing chart-based AUV monitoring interface with the MIRIAM interface (Figure \ref{fig:miriam2vehchatfilled}) has already been shown to improve situation awareness for experienced operators in a study with simulated AUV missions \cite{Robb2018}.

%% 2) Description of how the comms relay work and what it does (Jonatan?)
The complete system consists of multiple platforms, where one is dedicated to act as a Communication Relay (CR). This vehicle is on the surface, enabling it to use both underwater acoustic communication and Radio Frequency (RF) technologies such as WiFi. The CR vehicle extends the range of the AUV signal to the C2 centre, by multi-hop and medium change. This enables monitoring and control that would otherwise not have been possible. 

%DR- Not sure if we can describe this as an experiment
%We describe the results of the experiment performed at recent sea trials. 
We describe the results of recent sea trials. 
These highlight the benefits of using this combined system for this type of application. The combination of the MIRIAM interface along with a communications relay system in a Sonobot USV allowed an operator situated in a land based control room to chat in natural language with a remote IVER-3 AUV, which had been launched from a boat by a separate team.

In the sections below, we describe details of the Neptune Autonomy Framework which allows a high degree of autonomy among teams of AUVs, the MIRIAM natural language interface, a system of relayed communications and the results of sea trials of these components integrated together with a live IVER-3 AUV performing a survey with a Sonobot (Figure \ref{fig:sonobot}) taking the role of a communication relay. 
%%%%%%%%%%%%%%%%%%%%%%%%%%%%%%
%The Neptune Autonomy Framework
%%%%%%%%%%%%%%%%%%%%%%%%%%%%%%
\section{The Neptune Autonomy Framework}
%\textbf{WE NEED TO REWRITE AND OR AUGMENT THIS FROM THE ACCEPTED ABSTRACT}
%HH fix the citations %DR fixed
The Neptune Autonomy Framework from SeeByte makes use of techniques described in \cite{Miguelanez11, Petillot09}  to enable the cooperation of multiple unmanned autonomous vehicles. It allows the planning of missions by defining a set of objectives e.g. areas to be surveyed by patrolling in a search pattern while collecting sensor data or specific items of interest, to be reacquired by revisiting a location and following some reacquisition behaviour suitable for the type of goal and the available sensors. Once the objectives are defined, these are input and a rehearsal track calculated for each vehicle allocated to the mission. This is displayed on the map area of SeeByte's SeeTrack-Neptune user interface (Figures \ref{fig:miriam2vehchatfilled} and \ref{fig:3parscreenlive}) indicating the provisional solution for completing the mission objectives. The planned objectives are then uploaded to the vehicles and they can be launched to perform their tasks autonomously. 
%%DR I am adding here about autonomous behaviour to later link to benefits of a conversational assistant with explanation capability
Although it is possible for operators planning the mission to provisionally assign particular objectives to particular vehicles before uploading the mission, the autonomy framework may well result in vehicles undertaking the objectives differently once the mission is underway. This is due to the framework optimizing based on environmental conditions such as currents and vehicle capabilities. This is one aspect of autonomy that can lead to uncertainty in operators over whether a mission is proceeding as planned and that the AUVs are behaving appropriately. An intelligent, easy-to-use interface, such as the one described here, can help in this regard. 

%An interface that can reassure the operator that the mission is 

 %and an interface which could explain unexpected but valid autonomous behaviour would be desirable. Indeed the MIRIAM prototype, aside from enabling factual queries, does offer access to explanations of some autonomous behaviour.

%%%%%%%%%%%%%%%%%%%%%%%%%%%%%%
%Natural Language Interface
%%%%%%%%%%%%%%%%%%%%%%%%%%%%%%
\section{Natural Language Interface}

\begin{figure}[t]
\centering
\includegraphics[width=0.94\linewidth]{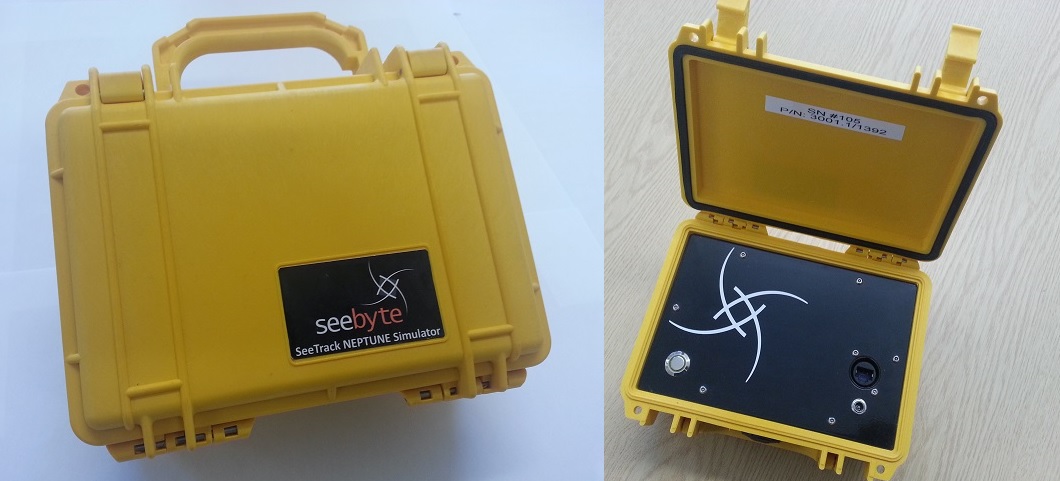}
\caption{The Neptune simulator used during MIRIAM development. It simulates either one or two AUVs running the Neptune Autonomy Framework (Figure \ref{fig:iveronpier}). Configuration files allow simulation of factors including the onset of various faults and accelerated battery drain due to adverse environment conditions.}
\label{fig:simulator}
\end{figure}

The MIRIAM natural language interface for autonomous systems was developed, building on the REGIME system for post-mission reporting \cite{hastieicmi2017trust}. MIRIAM integrates with the Neptune Autonomy Framework and has been fully tested in simulator (Figure \ref{fig:simulator}) but in the field trials described here, it is the first time that it had been tested with a real vehicle. 
The MIRIAM system connects to the Neptune Autonomy Framework via Neptune's Application Program Interface (API), polls Neptune and posts the mission and vehicle information in the mission database ready to be queried. MIRIAM accepts user queries; extracts the semantic content; formulates, executes and gathers the results of database queries; and, finally, constructs natural language replies for output. 

\begin{figure}[t]
\centering
\includegraphics[width=1.0\linewidth]{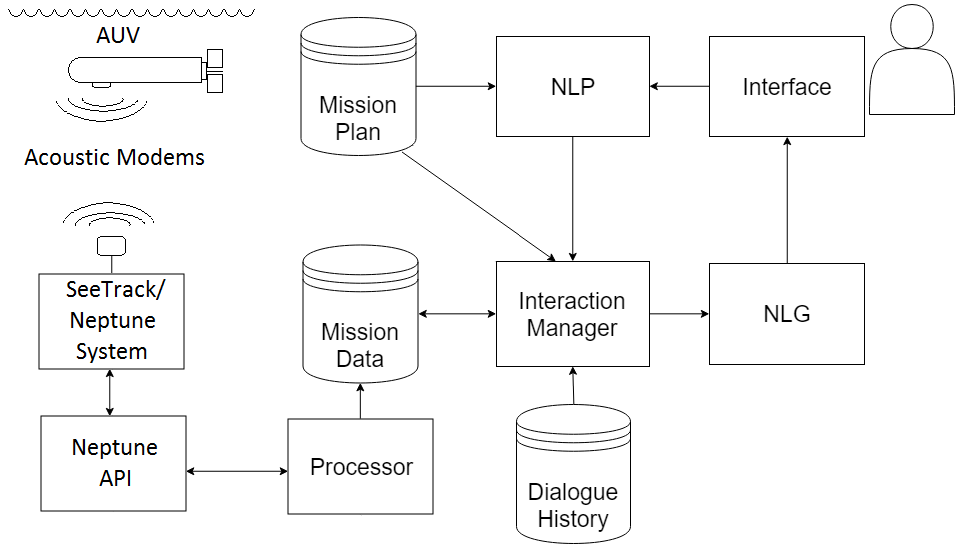}
\caption{The MIRIAM Natural Language Interface System Architecture. NLP/G is Natural Language Processing/Generation. }

\label{fig:sysarch}
\end{figure}

Figure \ref{fig:miriam2vehchatfilled} shows examples of the output while Figure \ref{fig:sysarch} shows the MIRIAM system architecture. Earlier prototypes enabled querying of vehicle and mission data such as vehicle speed and mission objectives. They also output notifications of important events such as completion of objectives and the occurrence of vehicle faults \cite{Hastie17demo}. The use of the natural language interface alongside the SeeTrack chart-based interface has been shown to be associated with improved situation awareness in operators \cite{Robb2018}. The recent MIRIAM prototype can also give explanations of some vehicle behaviours taking into account the current mission context \cite{Hastie2018Demo}. The explanations are based on an interpretable model of autonomy \cite{GarciaetalINLG18}. That model is constructed through having an expert ‘speak aloud’ to provide rationalization of the autonomous behaviours while watching videos of missions on the SeeTrack software. This method has the advantage of being agnostic to the model of autonomy and could be used to describe rule-based autonomous behaviours but also complex deep learning models. The resulting explanation mechanism provides two main types of explanation: "Why" and "Why not". These help operators to understand system capabilities and help establish appropriate levels of trust \cite{Lim2009}. 

MIRIAM can accept keyboard and voice input and outputs text and (optionally) speech. The natural language interaction it affords makes the vehicle and mission information highly accessible. This is an advantage in a remote, C2 centre where operators may not be entirely familiar with all the details, capabilities, and likely behaviours of every autonomous asset in a complex cooperative mission.

%%%%%%%%%%%%%%%%%%%%%%%%%%%%%%
%Relaying Communications using Multiple Vehicles
%%%%%%%%%%%%%%%%%%%%%%%%%%%%%%
\section{Relaying Communications using Multiple Vehicles}

%Acoustic communication is, to date, the only usable medium affording a long range means of communication in water. The acoustic method suffices although unreliable in nature as it suffers from several potential drawbacks such as low bandwidth, the characteristics of the acoustic channel, and package collision to mention a few. The range for a typical commercial acoustic modem is limited to several kilometres in good conditions. This limits the operational area, if a direct communication link from a C2 centre to an AUV is desired. 
Acoustic communication is, to date, the only usable medium affording a long range means of communication in water. While the acoustic method does work, it is unreliable in nature as it suffers from several potential drawbacks such as low bandwidth, the characteristics of the acoustic channel, and package collision. The range for a typical commercial acoustic modem is limited to several kilometres in good conditions. This limits the operational area, if a direct communication link from a C2 centre to an AUV is desired. 

An alternative approach
%DR - maybe we should not claim multplicity without enumerating or listing explicitly here?
%, with multiple benefits, 
is to deploy a communication relay (CR). A CR could be a static buoy or a moving platform, such as a USV. Both can extend the operational area of the C2 centre and the AUV, if a real-time communication link \cite{Vasilijevic2017} between these is required or desired. This can be done by enabling the USV and the C2 centre to use RF communication, which can relay the communication both ways. Another benefit is that spreading out possible CRs gives the possibility to cover more of the acoustic channel \cite{Freitag2000}, thereby increasing the possibility to receive the messages. In addition, the use of mobile, autonomous CRs, such as the EvoLogics Sonobot, adds the %possibility for smart behaviours to follow track and follow the AUVs \cite{Willners2018},
opportunity for smart behaviours to automatically track and follow the AUVs \cite{Willners2018},
further extending the operational area. 
%DR %JW (MIGHT NEED TO GET ANOTHER CITATION). 
%HH Possibility--> Likelihood?
%DR I think it is    Opportunity

\begin{figure}[t]
    \includegraphics[width=0.45\textwidth]{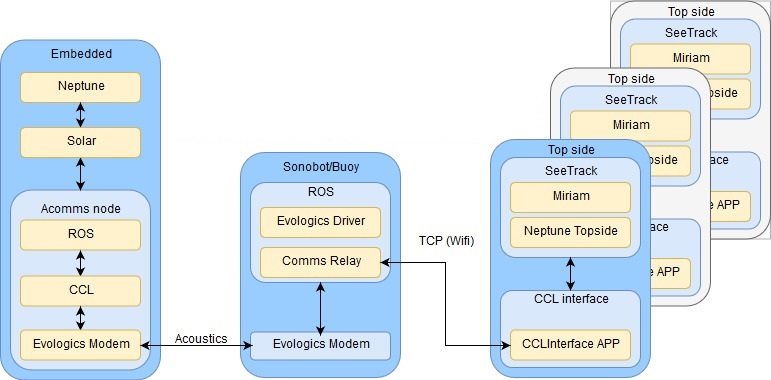}
\caption{Communication Relay ROS node, relaying acoustic communication to and from WiFi or other electromagnetic signals.}

\label{fig:commsrelay}
\end{figure}

The architecture for the system in the described scenario is displayed in Figure \ref{fig:commsrelay}.  The CR is implemented as a Robotic Operating System (ROS) \cite{Quigley2009} node. The node has two objectives, handling the relay of data and to act as a TCP server. For data, the node will listen to a topic where all incoming acoustic messages are %HH acoustics or acoustic? %DR acoustic
published, repack the message and send it to the TCP server. It will also listen to what is incoming to the TCP server, pack this to an acoustic message and publish this to be handled by the drivers for the acoustic modem. For the TCP server, any other machine, robot or other platform can connect to the node. It will relay all the information incoming from the acoustic communication to all connected clients.  
%HH incoming from the acoustic --> acoustic communication-- is that correct? or just "acoustics"
%DR I think "acoustic communication" is meant. 

\begin{figure}[t]
\centering
\includegraphics[width=0.94\linewidth]{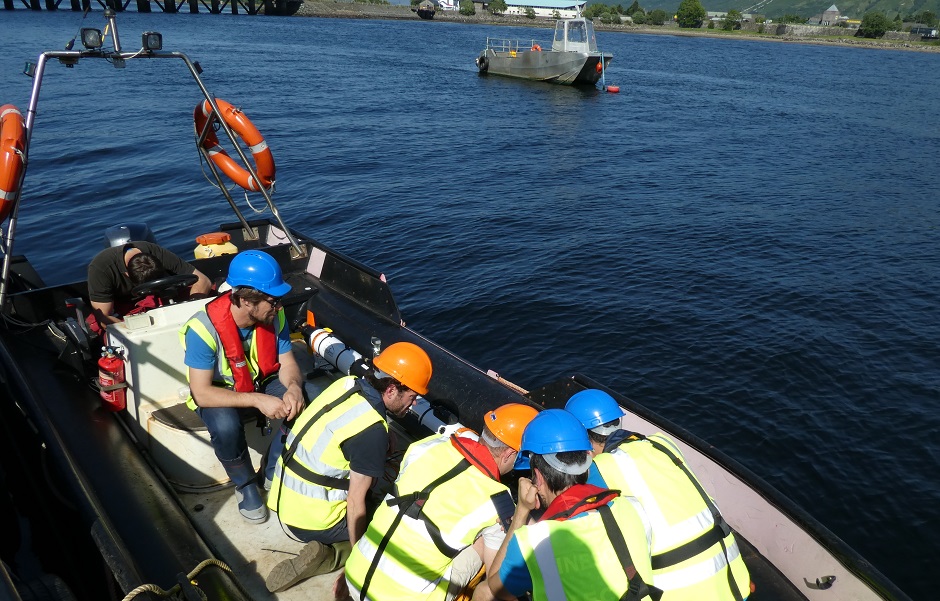}
\caption{Checking ACOMMS at the quayside with the IVER-3 AUV in the boat before launching the vehicle and starting the mission.}

\label{fig:quayside}
\end{figure}

%%%%%%%%%%%%%%%%%%%%%%%%%%%%%%
%Integration of Neptune, Chatbot and Relay Communications with an IVER AUV in Sea Trials
%%%%%%%%%%%%%%%%%%%%%%%%%%%%%%
\section{Integration of Neptune, MIRIAM and Relay Communications with an IVER-3 AUV in Sea Trials }

%\textit{The MIRIAM chatbot and the relayed acoustic communications were tested in a scottish sea loch integrated with an IVER-3 AUV and a Sonobot ASV at sea trials in a Scottish sea loch. The final paper will describe the steps taken and challenges encountered during the integration process, and discussion of future developments and implications.}

The sea trials were performed in a Scottish sea loch (a deep water sea inlet). The test scenario involved an AUV performing a seabed survey, while a surface vehicle was relaying acoustic communication between the AUV and the C2 centre.
%DR - OceanServer is the maker of the IVER-3 isn't it?
%The AUV used was an IVER-3 from Oceanscan with a computer running Neptune as a backseat driver. 
The AUV used was an IVER-3 from OceanServer with a computer running Neptune as a backseat driver. The surface vehicle was an EvoLogics Sonobot. Both vehicles were equipped with EvoLogics S2CR 18/34 OEM Acoustic modems. The C2 centre was located in a different location, without having direct access to an acoustic modem. From the C2 centre the mission could be started with a command relayed to be sent acoustically to the AUV. 
%DR - we name SeeByte earlier and later
%The sea bed survey mission was monitored with the SeeByte Ltd. SeeTrack graphical user interface and queried live using the MIRIAM chatbot. 
The seabed survey mission was monitored with the SeeTrack graphical user interface (by SeeByte Ltd.) and queried live using the MIRIAM interface. 

%DR Jonatan please edit my draft as you wish here 
%The integration of these elements was carried out in 6 days of trials over a period of six weeks. The integrated prototype system was demonstrated on the final day with an iver-3 sea bed survey mission being monitored with the SeeByte Ltd. SeeTrack chart display and queried live using the MIRIAM chatbot located remotely on-shore in an office acting as C2 centre. 

%JSW: I am not completely sure how to add this part, I see the point, but not exactly what it is aiming to contribute to. Please feel free to change however you like.
To reach this stage, a total of six days of trials were carried out, spread out over roughly six weeks. The trials proceeded in the following stages:
\begin{enumerate}%[label=\alph*]
\item First integration with EvoLogics modem. Drivers for modems were installed and tested on the AUV. In this step, the C2 computer was connected through serial communication to a Raspberry Pi running a ROS node for the modem drivers.
\item A Raspberry Pi running the ROS-node and a communication relay node was integrated on the Sonobot.
\item Testing with the MIRIAM software.

\end{enumerate}

\subsection{Apparatus}
%\begin{figure}[t]
%    \includegraphics[width=0.9\linewidth]{images/IVER_on_Pier-ResCrp.jpg}
%\caption{The IVER-3 AUV, sitting on its stand on the pier at the trials facility, being washed after recovery from a mission. }
%\label{fig:iveronpier}
%\end{figure}

\begin{figure}[t]
\centering
\includegraphics[width=0.94\linewidth]{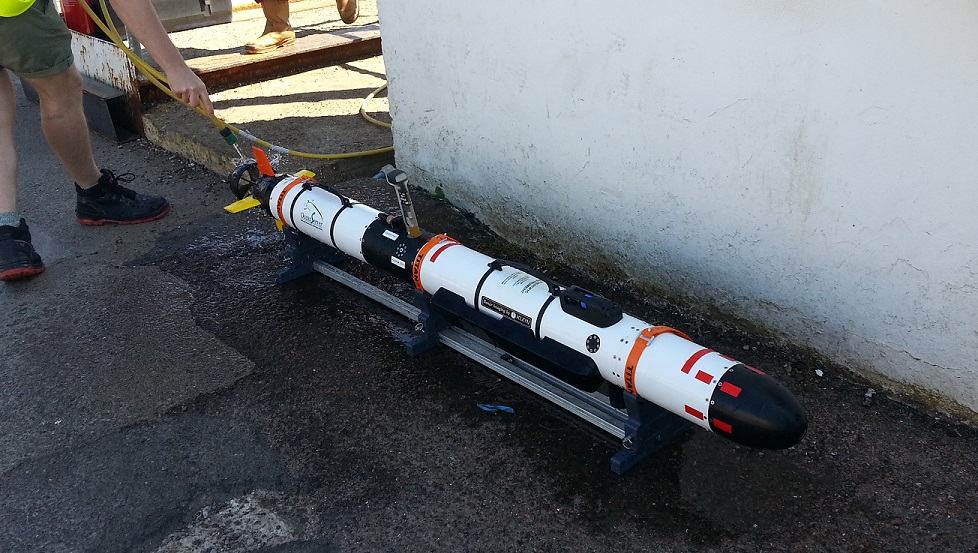}
\caption{The IVER-3 AUV, sitting on its stand on the pier at the trials facility, being washed after recovery from a mission. }

\label{fig:iveronpier}
\end{figure}

%DR-sonobotcarried- Jonatan please check and edit this caption
\begin{figure}[t]
\centering
\includegraphics[width=0.94\linewidth]{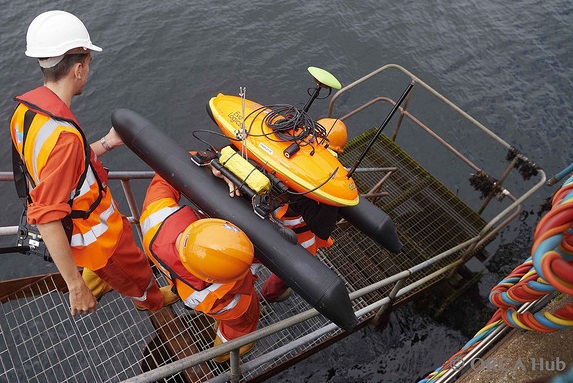}
\caption{The Sonobot USV being carried to a launching position below the pier at the test facility. Visible are its two electro-magnetic antennae masts and its S2CR 18/34 acoustic modem head (which is hung below the surface when deployed afloat). The vehicle's port side is facing the viewer. The yellow box fixed above the port side flotation hull contains the additional Raspberry Pi on-board computer and electronics for the CR.}
\label{fig:sonobotcarried}
\end{figure}

\subsubsection{IVER-3 Autonomous Underwater Vehicle}

The OceanServer IVER-3 (Figures \ref{fig:iveronpier} and \ref{fig:quayside}) is 2.1m in length, 0.14m in diameter, has endurance of 8-14 hours at 2.5 knots and speed of 1 to 4 knots. Its main sensor is high resolution side scan sonar. It has wireless and acoustic communications and GPS and Doppler velocity log (DVL), including depth sensor and compass, for navigation. On its embedded backseat computer, it runs the Neptune Autonomy Framework.

\subsubsection{S2CR 18/34 Underwater Acoustic Modem}
The acoustic modems used are produced by EvoLogics. The modems were configured to have a maximum range of 3500m and are able to send up to 13.9 kbit/s.
%DR Jonatan, please add any other hardware details here. They can be moved elswhere as appropriate later.

\subsubsection{Sonobot}
To carry the seabourn part of the CR, we used Sonobot (Figure \ref{fig:sonobot}), a USV from EvoLogics. It has its own control and planning software. It runs with the DUNE autonomy framework and is planned and controlled with Neptus. This particular vehicle has custom integrated S2CR 18/34 acoustic modems and an additional computer to handle relaying of communication. The CR system is not connected to DUNE.

\subsubsection{Laptop computers}
Two laptop computers running Microsoft Windows 10 operating system were used for control and monitoring of the two autonomous vehicles (IVER-3 and Sonobot). Both carried installations of Seebyte Ltd's SeeTrack software, which includes the Neptune Autonomy Framework. One laptop was used afloat in the small boat (a rib) used for launching and retrieval of the IVER-3 AUV. This laptop was also used to plan and upload the short survey mission to the IVER-3. The second laptop was used on-shore and, in addition to SeeTrack, was equipped with the DUNE and Neptus autonomy and control software for the Sonobot and also the MIRIAM software to query the IVER-3 through the Neptune API.

\subsubsection{WiFi Relay Equipment}
%DR-WiFi-Relay- Jonatan please check and edit this caption
In addition to the WiFi capability added to the Sonobot to enable the afloat end of the CR, equipment was assembled to handle on-shore relay. A WiFi antenna mast with booster was used to relay the received WiFi signal allowing the Sonobot relay USV to be further from the control centre than normal Wifi range would allow.

\begin{figure}[t]
\centering
    \includegraphics[width=0.94\linewidth]{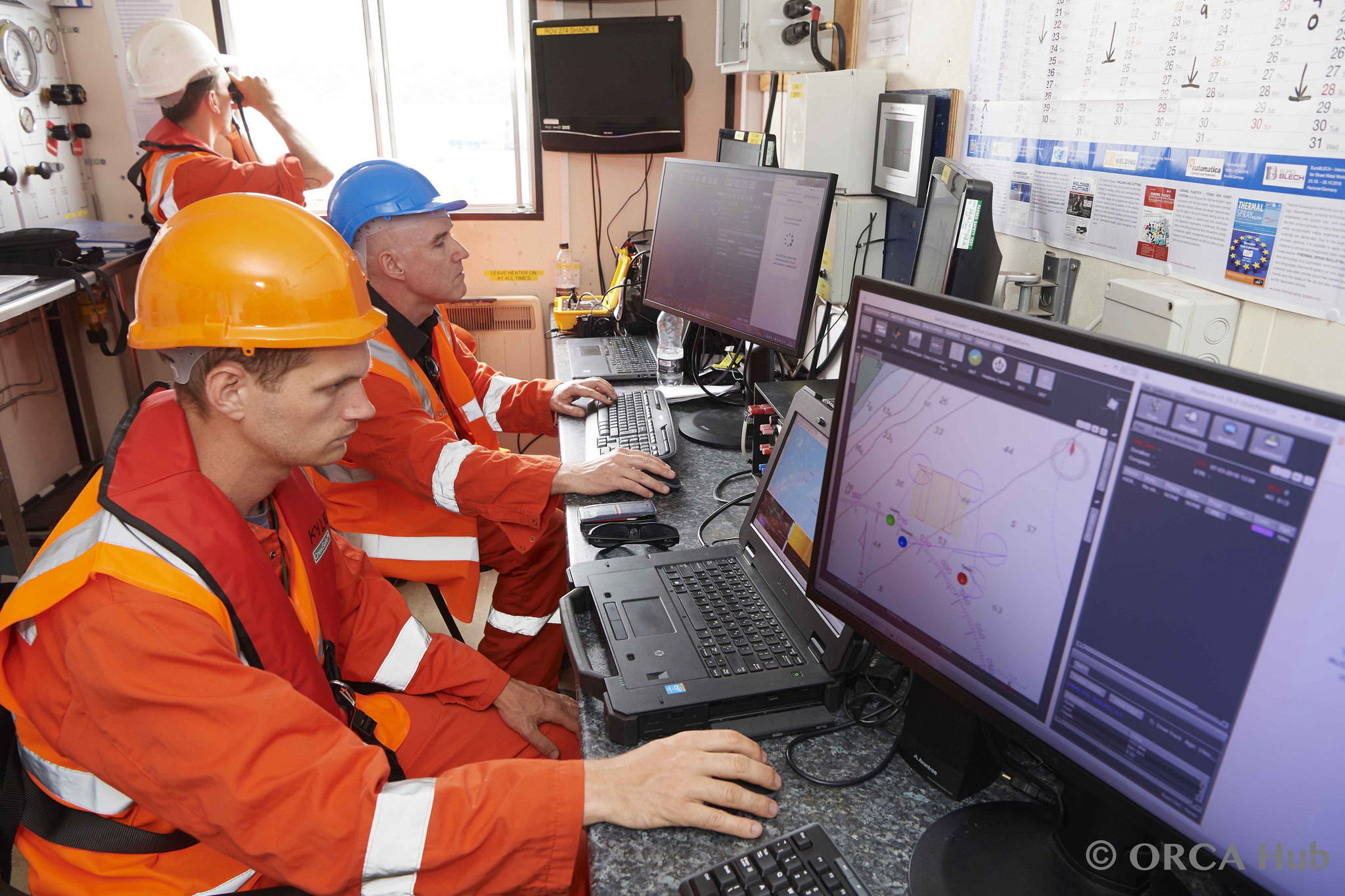}
\caption{On-shore control room. An operator controls the communications relay Sonobot USV separately using dedicated software. While the IVER-3 AUV mission can be monitored on the SeeTrack/Neptune/MIRIAM interfaces.}

\label{fig:controlroom}
\end{figure}

\begin{figure}[t]
\centering
\includegraphics[width=0.94\linewidth]{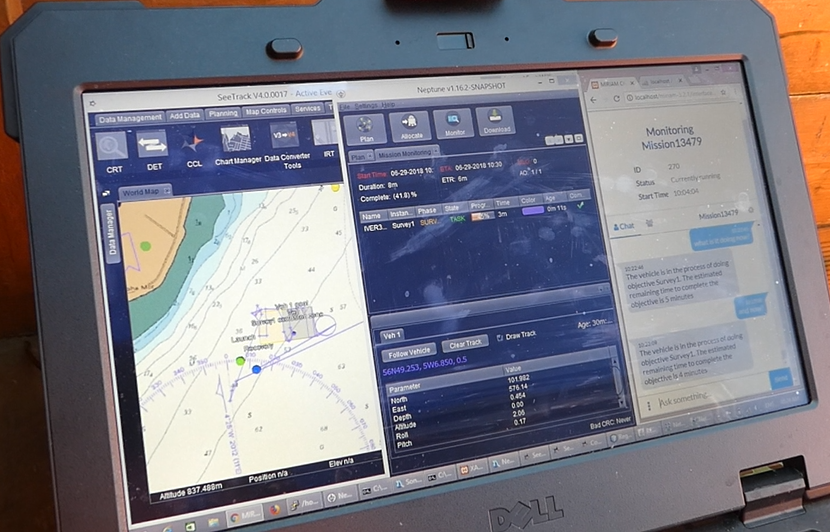}
\caption{A live mission monitored on-shore on a laptop. On the left side of screen is the SeeTrack map display showing the IVER-3 vehicle track with a simple mission consisting of a launch point, small survey area, and recovery point. In the middle of the screen is the Neptune window presenting mission and vehicle data in tabular form. The right hand part of the screen is the MIRIAM chat window for natural language interaction about the mission progress and vehicle state.}

\label{fig:3parscreenlive}
\end{figure}
%%%%%%%%%%%%%%%%%%%%%%%%%%%%%%
%Outcome of the Trials and Future Work
%%%%%%%%%%%%%%%%%%%%%%%%%%%%%%
\section{Outcome of the Trials and Future Work}

\subsection{The MIRIAM software}
%Overall success
Overall the trials were a success culminating in missions where the combined MIRIAM interface and SeeTrack interface was used to monitor and query the IVER-3 and the mission status (Figure \ref{fig:3parscreenlive}).
%Items for improvement
The trials exposed some practical issues in the MIRIAM software. Some of these relate to how MIRIAM was developed and tested in simulation rather than with a real vehicle. For example, in asset discovery it was found that the simulator behaviour was simulating AUVs as if they were always available on WiFi. However, this is not the case with real vehicles. Thus the asset discovery code in MIRIAM needs to be developed to handle intermittent asset discovery as occurs when vehicles are only available through acoustic communications.

%with a Neptune compatible AUV simulator while in the trials operations were with an IVER-3 AUV. 

%The main aspects which require addressing in the future are:
%HH we want to be careful we don't include our dirty laundry here.
%DR Leaving out
%\begin{enumerate}[label=\alph*]
%Firstly, the sequence of steps required in starting up the whole system including the MIRIAM prototype was a long one. E.g. MIRIAM alone requires a web server, database server, chatbot application and back-end application all to be launched (and this does not include those steps required to get the IVER, Neptune, SeeTrack and the communications relays up and running). This sequence requires automation in the future.

There were differences between the formats of the IVER-3 vehicle messages as presented in the Neptune API (to which MIRIAM connects) compared to those from the simulated vehicles. This did not affect the mission plan and objectives data, which were all relayed correctly by MIRIAM, but did affect the GPS location of the vehicle.

%However, while some of the vehicle status data was able to be successfully queried in natural language (e.g. battery level), some was not (e.g. GPS location). The MIRIAM software will need to be adapted to be flexible in how it gathers data from the API and posts it into the mission database for MIRIAM to access.

%\end{enumerate}
%Discussion of future work 
%on scalability/extensibility
One of the main  benefits of a conversational agent such as MIRIAM is in uniting data from multiple vehicles in a single interface. The latter two practical problems above point to the general need for adaptable proactive APIs or drivers to be positioned between the managerial/monitoring system and the autonomous vehicles' own system APIs. These would ideally allow intelligent identification of the appropriate fields for capture in any incoming data stream perhaps based both on expected data formats and content.

%on issuing commands throug chat
The MIRIAM prototype, as deployed in the trials, enables only information seeking by the operator. Commands such as \textit{mission start} and \textit{abort to recovery point} are currently issued through the SeeTrack interface. Future developments include enabling the issuing of commands, such as in-mission goal change, through the MIRIAM interface by voice or keyboard in natural language.

%DR Maybe Jonatan, you can add here?
\subsection{The Relayed Communications}
The usefulness of relayed communication was shown on several occasions during the trials, including one where the C2 centre was located in a place with no easy access to deploy its own acoustic modem. The CR node is light weight, making it able to run on multiple platforms. This would enable over the horizon communication and coverage of larger areas through the use of airborne vehicles, or multi-hop between surface vehicles.

%Use of dedicated Sonobot control laptop
The Sonobot, which carried the acoustic modem and relayed the communication to the C2 centre, was controlled using dedicated software on a separate laptop in the control room. In future all the vehicles involved could be running the Neptune Autonomy Framework and be able to follow mission plans created in SeeTrack including the CR vehicles. As any vehicle operating under Neptune can be compatible with the MIRIAM interface, this would make all the mission's assets accessible through MIRIAM. (The MIRIAM database is designed to allow additional vehicle types with different capability sets.)

\section{Conclusion}
Our goal in this work is to improve the effectiveness and efficiency of command and control (C2) of AUVs, which are increasingly tasked to cooperate together on complex missions. There is a requirement in hazardous and high-risk scenarios, such as those seen in the offshore energy sector \cite{orca}, to have personnel housed onshore and away from operations.  Addressing this requirement to have a distant C2 centre,  we have combined a natural language conversational interface with a Communication Relay system. MIRIAM connects to existing commercial C2 and autonomy software through an API, gathers the constantly updated mission and vehicle data, stores it in its database, accepts user queries, outputs replies and generates its own notifications of important events  in natural language. The CR system, implemented as a ROS node running on a computer on-board a USV, handles the relay of data, receiving ACCOMS, repackaging the messages, passing them to a TCP server for onward electromagnetic transmission and similarly in reverse.

We successfully demonstrated this system at trials in a Scottish sea loch using a) an OceanServer IVER-3 AUV as the vehicle to be controlled and monitored b) an EvoLogics Sonobot USV as the seabourn CR and c) Seebyte Seetrack-Neptune C2 software with the MIRIAM natural language interface in the on-shore C2 centre. 

%We also discussed future work and further developments to improve the prototype systems and better apply the concepts we have demonstrated.

%HH put on new column
%DR done

%%%%%%%%%%%%%%%%%%%%%%%%%%%%%%
%Acknowledgment
%%%%%%%%%%%%%%%%%%%%%%%%%%%%%% 
% EU H2020 Strongmar project (H2020-TWINN-2015, 692427).?
\section*{Acknowledgments}
%This work was supported by the EPSRC funded ORCA RAI-HUB (EP/R026173/1)
This work was supported by the EPSRC funded ORCA Hub (EP/R026173/1). The MIRIAM interface development was also supported by UK MOD Dstl Defence and Security Accelerator Programme for "Revolutionise the human information relationship for Defence" ACC101949 and RAEng/Leverhulme Trust (Hastie/LTSRF1617/13/37). We also thank the management and staff of the Underwater Centre, Fort William, UK.

%%%%%%%%%%%%%%%%%%%%%%%%%%%%%%
%bibliography
%%%%%%%%%%%%%%%%%%%%%%%%%%%%%%
%\bibliographystyle{./IEEEtran}
%\bibliography{./IEEEabrv,./mybib-add-refs-here}

\bibliographystyle{bibtex/bst/IEEEtran}
%\bibliography{bibtex/bib/IEEEabrv,bibtex/bib/mybib-add-refs-here, bibtex/bib/bib-from-abstract}
%\bibliography{bibtex/bib/IEEEabrv,bibtex/bib/mybib-add-refs-here}
%\bibliography{bibtex/bib/IEEEabrv,bibtex/bib/START-HEREconference_041818}
\bibliography{START-HEREconference_041818}

\end{document}